\documentclass{aa}
\usepackage{graphicx}
\usepackage{txfonts}

\begin{document}

\title{Optical polarimetric observations of the microquasar
\object{LS~5039}\thanks{Based on observations made at the Complejo
Astron\'omico El Leoncito, which is operated under agreement between CONICET
and the National Universities of La Plata, C\'ordoba, and San Juan.}}

\subtitle{}

\author{J.~A. Combi\inst{1,2}
\and S.~A. Cellone\inst{3}
\and J. Mart\'{\i}\inst{1}
\and M. Rib\'o\inst{4}
\and I.~F. Mirabel\inst{4,5}
\and J. Casares\inst{6}
}
\institute{Departamento de F\'{\i}sica, Escuela Polit\'ecnica Superior, Universidad de Ja\'en, 
Virgen de la Cabeza 2, 23071 Ja\'en, Spain\\
\email{jcombi@ujaen.es; jmarti@ujaen.es}
\and Instituto Argentino de Radioastronom\'{\i}a, C.C.5, (1894) Villa Elisa, Buenos Aires, Argentina
\and Facultad de Ciencias Astron\'omicas y Geof\'{\i}sicas UNLP, Paseo del Bosque, B1900FWA La Plata, Argentina\\
\email{scellone@fcaglp.fcaglp.unlp.edu.ar}
\and Service d'Astrophysique, CEA Saclay, B\^at. 709, L'Orme des Merisiers, 91191 Gif-sur-Yvette, Cedex, France\\
\email{mribo@discovery.saclay.cea.fr; mirabel@discovery.saclay.cea.fr}
\and Instituto de Astronom\'{\i}a y F\'{\i}sica del Espacio, CONICET,
C.C.67, Suc. 28, 1428 Buenos Aires, Argentina
\and Instituto de Astrof\'{\i}sica de Canarias, 38200 La Laguna, Tenerife, Spain\\
\email{jcv@ll.iac.es}
}


\offprints{J.~A. Combi,\\ \email{jcombi@ujaen.es}}

\date{Received / Accepted}

\abstract{
We present the first optical polarimetric observations of the runaway
microquasar \object{LS~5039}. Our results reveal the presence of a large
amount ($\sim$5\%) of polarized emission towards this binary system. By
combining polarimetric and spectroscopic observations of some stars in the
field together with available statistical information on the galactic
interstellar polarization of the region, we have estimated and subtracted the
contribution of the interstellar polarization in this direction. As a result,
we obtain an intrinsic polarization of $\sim$3\% for the object, much higher 
than what would be expected from jet emission in the optical domain. We
suggest that the polarized light originates by electron Thomson scattering in
the stellar envelope of the companion star. This allows us to constrain the
size of the region where the polarized emission originates, as well as
estimating the scattering electronic density and the wind velocity at such
distance.
}
\maketitle

\section{Introduction} \label{introduction}

The high mass X-ray binary system \object{LS~5039} is one of the about sixteen
confirmed Galactic microquasars (Paredes et~al. \cite{paredes02}; Rib\'o
\cite{ribo02t}). Recent astrometric studies carried out by Rib\'o et~al.
(\cite{ribo02}), show that it is a runaway system escaping from the Galactic
plane with a systemic velocity of $\sim$150~km~s$^{-1}$.


\object{LS~5039} is a bright $V$$\sim$11.2 star with an ON6.5V((f)) spectral
type (McSwain et~al. \cite{mcswain04}), located at a distance of $2.9 \pm
0.3$~kpc (Rib\'o et~al. \cite{ribo02}). The binary system has a short orbital
period of $P_{\rm orb}=4.4267 \pm 0.0005$ days and a high eccentricity of
$e=0.48 \pm 0.06$ (McSwain et~al. \cite{mcswain04}). At radio frequencies, the
source has a non-thermal spectral index with moderate, but not periodic,
variability. No strong radio outbursts have been ever detected (Mart\'{\i}
et~al. \cite{marti98}; Rib\'o et~al. \cite{ribo99}; Rib\'o
\cite{ribo02t}).

In X-rays, the source presents a hard spectrum up to 30~keV, but no pulsed nor
periodic emission has been detected (Rib\'o et~al. \cite{ribo99}; Reig et~al.
\cite{reig03}). The possibility that \object{LS~5039} is a $\gamma$-ray
emitter was suggested by Paredes et~al. (\cite{paredes00}), who proposed its
association with the unidentified $\gamma$-ray source
\object{3EG~J1824$-$1514} (Hartman et~al. \cite{hartman99}). In this context,
the $\gamma$-rays would be produced by inverse Compton upscattering of
lower-energy photons by the non-thermal relativistic $e^{-}$ population
responsible of the radio emission. Based on wind accretion models McSwain
et~al. (\cite{mcswain04}) suggest a neutron star as the compact object in the
binary system, which would have an inclination angle between 40\degr\ and
60\degr.

Although several physical parameters of the system are known, there is not yet
a clear picture allowing to explain the broadband emission detected from it.
Therefore, new observations that help to constrain the properties of the
system are needed.

Optical polarimetric observations are a useful tool in the study of intrinsic
polarization of X-ray binaries, since they can give us important information
about their physical and geometrical properties. This kind of studies have
been carried out by Scaltriti et~al. (\cite{scaltriti97}) for the microquasar
\object{GRO~J1655$-$40}, where they found a significant amount of intrinsic
linear polarization ($\sim$3\%) in the $VRI$ bands, with the polarization
direction being parallel to the accretion disk plane. In a following paper on
this object, Gliozzi et~al. (\cite{gliozzi98}) detected oscillations in the
polarization consistent with the orbital period.

The intrinsic polarization in X-ray binary systems can result from nonthermal
emission processes due to the presence of high-energy electrons in the
relativistic jets or, alternatively, by electron Thomson scattering due to an
extended plasma above an accretion disk or a stellar envelope. Moreover, this
intrinsic polarization could change with the binary phase, due to the
polarizing mechanism, to the distribution and physical state of the material,
or by geometric factors such as the inclination angle.

The main problem in the effective identification of intrinsic polarization of
a source at optical frequencies is the contamination produced by the Galactic
interstellar medium (ISM) polarization. This component is thought to arise due
to extinction by anisotropic dust particles preferentially oriented by the
Galactic magnetic field. At present, the most complete study of Galactic
starlight polarization has been carried out by Fosalba et~al.
(\cite{fosalba02}). They presented a statistical analysis of the Galactic ISM
polarization using $\sim$5500 stars and found that the interstellar grains are
not fully aligned with the Galactic magnetic field, a fact that was
interpreted as the effect of a large random component of the field. However,
it is important to stress that an entire reconstruction of the
three-dimensional magnetic orientations requires additional radio,
submillimeter and infrared data.

In this paper, we report the first optical polarimetry of \object{LS~5039}. We
describe the observations in Sect.~\ref{observations}, present the obtained
results in Sect.~\ref{results}, and we discuss on the origin of the detected
polarization and state our conclusions in Sect.~\ref{discussion}.

\section{Optical observations}
\label{observations}

\subsection{Polarimetry}

Polarimetric observations of \object{LS~5039} were obtained at the 2.15\,m
``Jorge Sahade'' telescope, CASLEO, Argentina, using a two-channel
polarimeter. This instrument is a rotating plate polarimeter with a Wollaston
prism that splits the incident light beam into two components, each one
directed to a different photomultiplier (Mart\'{\i}nez et~al.
\cite{martinez90}). Its optical design is based on VATPOL, the Vatican
Observatory polarimeter (Magalh\~aes et~al. \cite{magalhaes84}).

The observations spanned four consecutive nights, roughly centered at the new
moon, from 2003 June 28 to July 1 (UT dates). Weather conditions were fairly
good for most of the observing time, except for thick cirrus during the second
night and thin cirrus during a few hours at the beginning of the first and
third nights. The target was repeatedly observed through $V$ and $I$ filters,
closely matching the Johnson-Cousins system, with individual exposure times of
300--360~s; this yielded between 8 and 12 individual data points each night in
each band. Sky integrations were regularly obtained between science exposures.
A 11.3\arcsec\ diaphragm was used for all the observations.

\begin{figure}[t!]
\resizebox{\hsize}{!}{\includegraphics{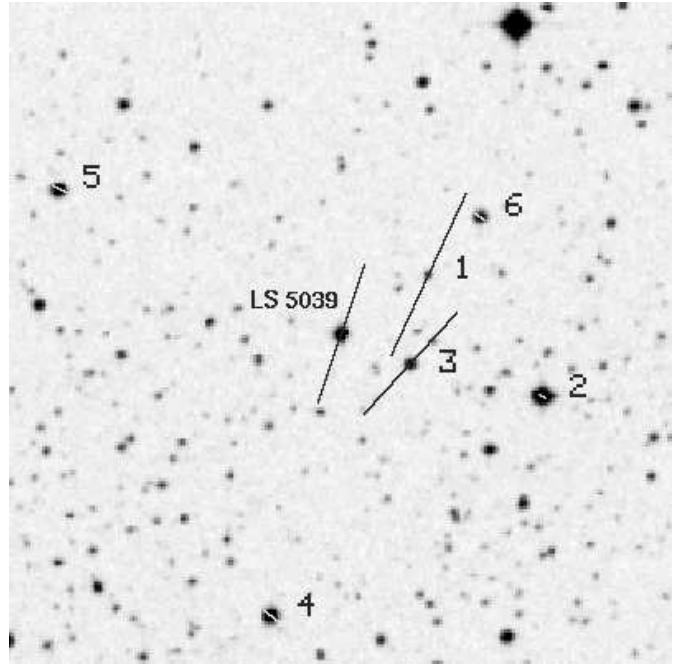}}
\caption[]{DSS 7\arcsec$\times$7\arcsec\ image centered on \object{LS~5039}, showing 
the target and the six comparison stars. Polarization vectors in the $V$ band have been 
overlaid. North is up and East is to the left.}
\label{fig:dss}
\end{figure}

Six field stars, with apparent magnitudes bracketing that of \object{LS~5039}
and within $\sim$3.5\arcmin\ from its position (see Fig.~\ref{fig:dss}), were
also observed in order to estimate the foreground polarization. Both polarized
and unpolarized stars from the lists of Turnshek et~al. (\cite{turnshek90})
were also observed each night, in order to evaluate the instrumental
polarization (which was found to be practically zero) and to correct the
measured polarization angles to the equatorial standard system. We were able
to determine this zero point correction to an accuracy of $\pm$5\degr.

Since both the ordinary and the extraordinary rays are measured almost
simultaneously, polarimetric observations should be barely affected by
non-photometric conditions. However, large and/or rapid variations in the sky
polarization (e.g., when the Moon is rising or setting) are expected to
introduce systematic errors in the measured polarization, because the sky
conditions may have changed between the object and sky measurements. To
prevent against this source of error, we interpolated the sky brightness and
polarization (in the Stokes $U-Q$ plane) to the same instant of each object
observation.

As a further check, we plotted the magnitudes and Stokes parameters of both
object and sky as a function of time, in order to verify whether or not any
change in the observed target polarization may have been caused by a sudden
change in the sky polarization and/or transparency. This allowed us to reject
some suspicious data points, as will be explained below.

\subsection{Spectroscopy}

Spectroscopic observations for the most polarized ($\sim$5\%) comparison stars
were conducted on 2003 October 13 with the 2.56~m Nordic Optical Telescope
(NOT) in the island of La Palma (Spain). The instrument used was the Andalucia
Faint Object Spectrograph and Camera (ALFOSC) combined with Grism~6 and the
CCD detector EEV42-40 (2K$\times$2K). The CCD frames were processed using
standard tasks in the IRAF package, including bias subtraction, flat-fielding,
optimal extraction and wavelength calibration based on arc frames taken with a
helium lamp.



\section{Results}
\label{results}


The polarization and position angle measurements in the $V$ and $I$ bands of
the six comparison stars, and average values for \object{LS~5039}, are quoted
in Table~\ref{table:pol}. We show in Fig.~\ref{fig:dss} a Digitized Sky Survey
(DSS) image of the target field with polarization vectors in the $V$ band
overlaid onto \object{LS~5039} and the six comparison stars.


It is evident that, at least, two different polarization patterns are produced
by interstellar dust along the studied direction. Stars \#2, \#4, \#5, and \#6
show polarization vectors with small amplitudes ($\langle P_V
\rangle$$\sim$0.3--0.6\%), roughly aligned in the NE-SW direction ($\langle
\theta_V \rangle$$\sim$60\degr). On the other hand, light from stars \#1 and
\#3 is polarized almost at an orthogonal direction ($\langle \theta_V
\rangle$$\sim$ 135--155\degr), with a significantly higher percentage
($\langle P_V \rangle$$\sim$5--6\%); hence, these two stars have polarization
properties very similar to that of \object{LS~5039}.

\begin{table}
\begin{center}
\caption[]{Average polarization and position angle measurements of the comparison stars and \object{LS~5039}.}
\label{table:pol}
\begin{tabular}{@{}c@{~~}c@{~~~}c@{~~~~~}c@{~~~}c@{}}
\hline \hline \noalign{\smallskip}
Object Name     & $\langle P_V \rangle$ & $\langle \theta_V \rangle$ & $\langle P_I \rangle$ & $\langle \theta_I \rangle$ \\
USNO-A2.0 0750- &  (\%)          & (\degr)        & (\%)          & (\degr) \\
\noalign{\smallskip} \hline \noalign{\smallskip}
13157416 (\#1)  & 6.36$\pm$0.41 &  155.2$\pm$1.8 & 4.75$\pm$0.11 &  155.9$\pm$0.7 \\
13153701 (\#2)  & 0.32$\pm$0.04 & ~~67.7$\pm$3.7 & 0.42$\pm$0.01 & ~~58.5$\pm$1.0 \\
13157934 (\#3)  & 4.92$\pm$1.00 &  137.5$\pm$5.8 & 5.67$\pm$0.08 &  162.3$\pm$0.4 \\
13162766 (\#4)  & 0.58$\pm$0.02 & ~~54.2$\pm$1.2 & 0.45$\pm$0.02 & ~~48.7$\pm$1.3 \\
13170022 (\#5)  & 0.51$\pm$0.04 & ~~65.2$\pm$2.0 & 0.47$\pm$0.03 & ~~57.2$\pm$1.7 \\
13155723 (\#6)  & 0.47$\pm$0.07 & ~~54.8$\pm$4.3 & 0.40$\pm$0.06 & ~~58.1$\pm$3.9 \\
\noalign{\smallskip} \hline \noalign{\smallskip}
\object{LS~5039} & 5.26$\pm$0.15 & 161.3$\pm$0.8 & 4.70$\pm$0.16 &  162.4$\pm$1.0 \\
\noalign{\smallskip} \hline
\end{tabular}
\end{center}
\end{table}

\begin{figure}[t!]
\resizebox{\hsize}{!}{\includegraphics{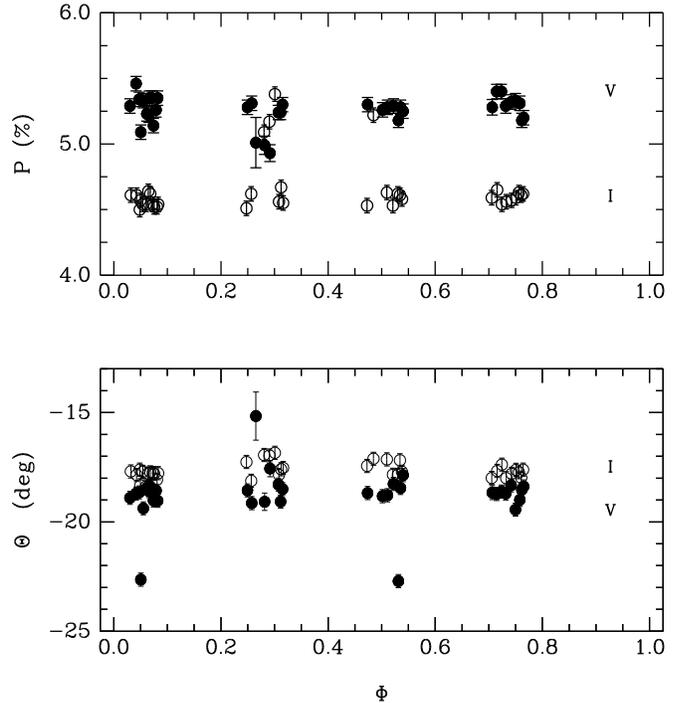}}
\caption[]{Polarization and position angle in the $V$ (filled circles) and $I$ (open circles)
bands as a function of the orbital phase for \object{LS~5039}.}
\label{fig:ls5039}
\end{figure}

We show in Fig.~\ref{fig:ls5039} the polarization measurements of
\object{LS~5039} in the $V$ (filled circles) and $I$ (open circles) bands as
function of the orbital phase ($t_0$=HJD\,2452756.49, $P_{\rm orb}$=4.4267~d).
During the first night, corresponding to $\Phi$$\simeq$0.05 in
Fig.~\ref{fig:ls5039}, the sky polarization remained fairly constant at
$\sim$1.5\%, both in $V$ and $I$. Since, in addition, its flux contribution
within the diaphragm was $\sim$4.5--5 mag lower than that from
\object{LS~5039}, no significant systematic errors from sky polarization are
expected. The $I$ band polarization shows no significant variations with time,
remaining nearly constant at $\langle P_I \rangle = 4.6$\%. In turn, the $V$
band polarization time-curve shows fluctuations at a $\la$3$\sigma$ level,
except for two points falling significantly below the mean. These data
correspond to brightenings in the light-curve, and the larger variation in
$P_V$ was accompanied by a $\sim$4\degr\ change in $\theta_V$. Although
instrumental errors as the sources for these variations cannot be absolutely
ruled out, there is no indication that any such error could have induced the
observed behaviour. However, this would contrast with the apparent photometric
stability of \object{LS~5039} within few hundredths of magnitude reported by
several authors (e.g. Mart\'{\i} et~al. \cite{marti04}).

On the second night ($\Phi$$\simeq$0.30), as stated above, we observed through
thick cirrus. As a consequence, the quality of these data is not as good, with
large fluctuations (up to $\sim$4\%) in the sky polarization; hence,
significant systematic errors could be affecting this data set. The first half
of the third night ($\Phi$$\simeq$0.50) was affected by thin cirrus, so we had
to discard a few suspicious data points. Finally, on the fourth night
($\Phi$$\simeq$0.75) we observed mostly under photometric conditions. The
object was constantly $\sim$4 mag brighter than the sky within the diaphragm,
and no rapid changes in the sky polarization were evident. No significant
systematic errors should therefore be expected.

Concerning the NOT spectroscopic observations, the sky conditions were
non-photometric and affected by clouds. Moreover, we could only acquire
spectra with the objects being at a significant air mass. Consequently, a
spectrum suitable for classification was only obtained for the brightest
comparison star \#3. After combining two 600~s exposures, the resulting
spectrum is consistent with an A0~Ib star taking into account the strong
narrow hydrogen absorption lines.

On the other hand, the apparent magnitudes of star \#3 are
$m_B$=13.92$\pm$0.04 and $m_V$=12.86$\pm$0.04, hence $B-V$=1.06$\pm$0.06
(Mart\'{\i} et~al. \cite{marti04}). For an A0 supergiant, the absolute
magnitude is $M_V$=$-$5.3$\pm$0.1 and the intrinsic color $(B-V)_0$=0.00
(Allen \cite{allen73}). Therefore, this comparison star has a color excess of
$E(B-V)$=1.06$\pm$0.06 magnitudes. Using the relationship
$A_V=(3.30+0.28(B-V)_{0}+0.04E_{B-V})\,E(B-V)$ (Schmidt-Kaler
\cite{schmidt82}), this corresponds to an interstellar extinction of
$A_V$=3.5$\pm$0.2 and a distance of 8.4$\pm$0.1~kpc.


\section{Discussion}
\label{discussion}

The observed polarization in astrophysical objects is usually the sum of two
components, the intrinsic one produced in the object and the one arising from 
the foreground ISM between the object and the observer. The high polarization 
measured towards \object{LS~5039} ($\sim$5\% in $V$ band) suggests that, at
least, part of it should have an intrinsic origin because it is very unusual
to be produced in the ISM up to a distance of 3~kpc. 

To deduce the intrinsic polarization of the source, $P_{\rm IN}$, it is
necessary to know the contribution of the ISM polarization, i.e. the value of
$P_{\rm ISM}$ and subtract it from the observed one. 


First of all, we assume that light emitted from stars \#1 to \#6 is not
polarized. The similar percentages and angles of polarization of stars \#2,
\#4, \#5 and \#6, suggest that they are being slightly affected by the ISM and
that they lie close to the Sun, possibly in the local spiral arm (distance
$<$500~pc). On the contrary, stars \#1 and 3\# have polarization properties
similar to that of \object{LS~5039}. 

As we have mentioned in previous sections, the main problem to determine the
intrinsic polarization of an object, is the absence of a complete study on the
Galactic interstellar polarization. However, we can obtain a rather crude
estimation of the intrinsic polarization of \object{LS~5039} by combining the
statistical study of galactic interstellar polarization carried out by Fosalba
et~al. (\cite{fosalba02}) and all the information gathered by us from
polarimetric and spectroscopic observations of the other stars in the field. 

Fosalba et~al. (\cite{fosalba02}), have found that the distribution of the
polarization degree as a function of distance grows linearly up to
$\sim$2~kpc, but beyond this distance the behavior of $P$(\%) is best fitted
by third-order polynomials up to a distance of $\sim$6~kpc (see their Fig.~3).
Adopting a distance of $\sim$2.9~kpc for \object{LS~5039} (Rib\'o et~al.
\cite{ribo02}) the polarization percentage as a function of the distance
results $\sim$2.3\%. From the equations that give the behavior of the
polarization degree with Galactic longitude and latitude, we obtain values for
$P$(\%)$\sim$0.8 and $P$(\%)$\sim$2, respectively. On the other hand, if we
assume that the contribution of the interstellar polarization increases
linearly with distance, we derive $P_{\rm ISM}$$\sim$2.5\% a value consistent
with those obtained from Fosalba et~al. (\cite{fosalba02}), for the set of
known stellar parameters of \object{LS~5039}. So, we suggest that the
intrinsic polarization of \object{LS~5039} is $\sim$3\%.

This value is also similar to the one obtained by Gliozzi et~al.
(\cite{gliozzi98}) for the microquasar \object{GRO~J1655$-$40}, which was
interpreted as the consequence of electron scattering by plasma above the
accretion disk.

The origin of the intrinsic polarized optical emission estimated for
\object{LS~5039} could be due to a nonthermal emission process or to electron
Thomson scattering. If the observed polarized flux is the result of
synchrotron emission originated from the relativistic plasma inside the jet,
then the radio electron populations should play an important role in the
generation of optical emission. However, we have extrapolated the energy 
distribution of the jet in \object{LS~5039} (Rib\'o \cite{ribo02t}) from 
radio to optical frequencies and find that it is at least 4 orders of
magnitude lower than the observed flux. Since the nonthermal emission process
is not enough to generate the polarized optical emission detected towards
\object{LS~5039}, the electron Thomson scattering appears as a suitable
alternative mechanism causing the observed polarized flux. Since there is no
clear evidence of the presence of a large accretion disk in \object{LS~5039},
the most likely possibility is that the polarized emission originates in the
stellar envelope of the companion star. Such scenario has been previously
proposed by Brown \& McLean (\cite{brown77}) to explain the intrinsic
polarization by Thomson scattering on free electrons in optically thin stellar
envelopes. An improved treatment is given in Brown et~al.
(\cite{brown78}), while other works on stellar scattering polarization can be
found in Cassinelli et~al. (\cite{cassinelli87}) and references therein. 

Considering the Brown \& McLean picture for the generation of intrinsic
polarization, we can estimate some physical and geometrical parameters for
\object{LS~5039}. Assuming that the plasma fills an ellipsoidal shell of any
thickness with a uniform density distribution $n_{0}$ of scattering electrons,
the expected polarization is given by:
\begin{equation}
P= \sigma_{T} n_{0} R (1-3\gamma) \sin^2 i
\end{equation}
where, $\sigma_{T}$ is the Thomson cross-section, $R$ is the radius of the
envelope, $i$ is the inclination angle of the equatorial plane and $\gamma$ is
a function of the ratio $a$ of the equatorial to polar radius. Assuming a
simple radial flow, the corresponding electron density of the envelope is:
\begin{equation}
n_{0} \sim \frac{\dot{M}_{\rm opt}}{4 \pi R^{2}
v_{\infty} \left[1- \frac{R_{\rm opt}}{R} \right]^{\beta} \mu m_{H}}
\end{equation}
Following McSwain et~al. (\cite{mcswain04}), suitable values for
\object{LS~5039} are a wind velocity at infinity
$v_{\infty}$=2440$\pm$190~km~s$^{-1}$, $\beta$=1.0, an optical star radius
$R_{\rm opt}$=8.5~$R_{\sun}$ and a wind mass loss $\dot{M}_{\rm
opt}$=4$\times$10$^{-8}$ $M_{\sun}$~yr$^{-1}$ (the last two quantities are
mean values). The other symbols have their usual meanings. Introducing Eq.~(1)
into Eq.~(2) and considering values of $\gamma$ of 0.1, 0.2 and 0.3
($P$(\%)$>$0) we estimate that the density of the scattering electrons is in
the range (0.2--1.3)$\times$10$^{12}$~cm$^{-3}$. From the above results we
conclude that the polarized emission originates mainly close to the star at a
distance between (8.51--8.6)~$R_{\sun}$, where the velocity of the stellar
wind is (3--20)~km~s$^{-1}$.

An interesting feature in the polarization of \object{LS~5039} is the detected
variability during the first and second nights. If the polarimetric
fluctuations observed on timescales of hours are real, then they might be
attributed to localized density enhancements (or blobs) propagating in the
general wind. The existence of these blobs has been introduced by Brown et~al.
(\cite{brown95}) to explain transient features seen in polarimetric light
curves of WR stars. Further simultaneous photometric and polarimetric
observations of the system are necessary to shed light on the origin of the
variable optical polarization.



\begin{acknowledgements}

We acknowledge Pablo Fosalba for useful discussions.
J.A.C. is a researcher of the programme {\em Ram\'on y Cajal} funded jointly
by the Spanish Ministerio de Ciencia y Tecnolog\'{\i}a and Universidad de
Ja\'en, and was also supported during this work by CONICET (under grant PEI 6384/03). 
J.A.C. is very grateful to staff of the Service d'Astrophysique (CEA Saclay)
where part of this research was carried out.
J.M. and M.R. acknowledge partial support by DGI of the Ministerio de Ciencia
y Tecnolog\'{\i}a (Spain) under grant AYA2001-3092, as well as partial support
by the European Regional Development Fund (ERDF/FEDER). 
J.M. is also supported by the Junta de Andaluc\'{\i}a (Spain) under project
FQM322. M.R. acknowledges support by a Marie Curie Fellowship of the European
Community programme Improving Human Potential under contract number
HPMF-CT-2002-02053. The Digitized Sky Survey was produced at the Space Telescope Science Institute
under U.S. Government grant NAG~W-2166.

\end{acknowledgements}


\begin{thebibliography}{}

\bibitem[1973]{allen73}
Allen, C.~W.
1973, Astrophysical Quantities, The Athlone Press Ltd., London

\bibitem[1977]{brown77}
Brown, J.~C., \& McLean, I.~S.
1977, A\&A, 57, 141

\bibitem[1978]{brown78}
Brown, J.~C., McLean, I.~S., \& Emslie, A.~G.
1978, A\&A, 68, 415

\bibitem[1995]{brown95}
Brown, J.~C., Richardson, L.~L., Antokhin, I., et~al.
1995, A\&A, 295, 725

\bibitem[1987]{cassinelli87}
Cassinelli, J.~P., Nordsieck, K.~H., \& Murison, M.~A.
1987, ApJ, 317, 290

\bibitem[2002]{fosalba02}
Fosalba, P., Lazarian, A., Prunet, S., \& Tauber, J.~A.
2002, ApJ, 564, 762

\bibitem[1998]{gliozzi98}
Gliozzi, M., Bodo, G., Ghisellini, G., Scaltriti, F., \& Trussoni, E.
1998, A\&A, 337, L39

\bibitem[1999]{hartman99}
Hartman, R.~C., Bertsch, D.~L., Bloom, S.~D., et~al.
1999, ApJS, 123, 79

\bibitem[1984]{magalhaes84}
Magalh\~aes, A.~M., Benedetti, E., \& Roland, E.~H.
1984, PASP, 96, 383

\bibitem[1998]{marti98}
Mart\'{\i}, J., Paredes, J.~M., \& Rib\'o, M.
1998, A\&A, 338, L71

\bibitem[2004]{marti04}
Mart\'{\i}, J., Luque-Escamilla, P., Garrido, J.~L., Paredes, J.~M., \& Zamanov, R.
2004, A\&A 418, 271

\bibitem[1990]{martinez90}
Mart\'{\i}nez, E., Aballay, J.~L., Mar\'un, A., \& Ruartes, H.
1990, Bol. Asoc. Arg. de Astronom\'{\i}a, 36, 342


\bibitem[2002]{mcswain02}
McSwain, M.~V., \& Gies, D.~R.
2002, ApJ, 568, L27

\bibitem[2004]{mcswain04}
 McSwain, M.~V., Gies, D.~R., Huang, W., et~al.
 2004, ApJ, 600, 927


\bibitem[2000]{paredes00}
Paredes, J.~M., Mart\'{\i}, J., Rib\'o, M., \& Massi, M.
2000, Science, 288, 2340

\bibitem[2002]{paredes02}
Paredes, J.~M., Rib\'o, M., Ros, E., Mart\'{\i}, J., \& Massi, M.
2002, A\&A, 393, L99

\bibitem[2003]{reig03}
Reig, P., Rib\'o, M., Paredes, J.~M., \& Mart\'{\i}, J.
2003, A\&A, 405, 285

\bibitem[2002]{ribo02t}
Rib\'o, M.
2002, PhD Thesis, Universitat de Barcelona

\bibitem[1999]{ribo99}
Rib\'o, M., Reig, P., Mart\'{\i}, J., \& Paredes, J.~M.
1999, A\&A, 347, 518

\bibitem[2002]{ribo02}
Rib\'o, M., Paredes, J.~M., Romero, G.~E., et~al.
2002, A\&A, 384, 954

\bibitem[1997]{scaltriti97}
Scaltriti, F., Ghisellini, G., Gliozzi, M., \& Trussoni, E.
1997, A\&A, 325, L29

\bibitem[1982]{schmidt82}
Schmidt-Kaler, Th.
1982, Landolt-B\"ornstein: Numerical Data and Functional Relationships in Science and Technology, New Series "Group 6 Astronomy and Astrophysics", Volume 2

\bibitem[1990]{turnshek90}
Turnshek, D.~A., Bohlin, R.~C., Williamson, R.~L., et~al.
1990, AJ, 99, 1243

\end{thebibliography}
\end{document}